\documentclass[sn-nature]{sn-jnl}
\usepackage{graphicx}%
\usepackage{multirow}%
\usepackage{amsmath,amssymb,amsfonts}%
\usepackage{standalone}
\usepackage{amsthm}%
\usepackage{mathrsfs}%
\usepackage[title]{appendix}%
\usepackage{xcolor}%
\usepackage{textcomp}%
\usepackage{manyfoot}%
\usepackage{booktabs}%
\usepackage{algorithm}%
\usepackage{algorithmicx}%
\usepackage{algpseudocode}%
\usepackage{listings}%
\usepackage{tikz}
\usepackage{pgfplots}
\pgfplotsset{compat=1.18}
\usepackage{tikz}
\usepackage{tikz}
\usepackage{pgfplots} 
\usepgfplotslibrary{groupplots} 
\usetikzlibrary{arrows.meta, positioning, shapes, patterns, calc, backgrounds, decorations.pathreplacing, shadows}
\usetikzlibrary{arrows.meta, positioning, shapes, patterns, calc, backgrounds, decorations.pathreplacing}
\usepackage{amsmath}
\usepackage{amssymb}
\usetikzlibrary{arrows.meta, positioning, shapes, patterns, calc}
\usepackage{caption} 

\theoremstyle{thmstyleone}%
\newtheorem{theorem}{Theorem}

\theoremstyle{thmstyletwo}%
\newtheorem{remark}{Remark}%

\theoremstyle{thmstylethree}%
\begin{document}


\title[LOB Dynamics in Matching Markets]{Limit Order Book Dynamics in Matching Markets: Microstructure, Spread, and Execution Slippage}

\author*[1]{\fnm{Yao} \sur{Wu}}\email{wuyao@westlake.edu.cn}
\affil[1]{
    \orgdiv{School of Engineering}, 
    \orgname{Westlake University}, 
    \orgaddress{
        \city{Hangzhou}, 
        \state{Zhejiang}, 
        \country{China}
    }
}

\abstract{
Conventional models of matching markets often assume that compensatory transfers (e.g., bride price) can clear markets by offsetting utility differentials. However, empirical evidence suggests that such transfers frequently fail to close the ``psychological gap'' between partners, leading to persistent inefficiencies. This paper proposes a \textbf{market microstructure framework} that formalizes mate selection not as a bargaining game, but as a \textbf{Limit Order Book (LOB) system} characterized by rigid bid--ask spreads.

We model individual preferences using a \textbf{Latent Preference State Matrix (LPSM)}, where the decision rule is driven by the structural spread between an agent's \textbf{internal ask price} (Unconditional Max) and the market's \textbf{best available bid} (Reachable Max). We formally demonstrate that this \textbf{Internal Preference Differential} ($\Delta V$) functions as a structural liquidity constraint that is \textbf{immune to linear monetary compensation}. 

\begin{equation}
\Delta V = V_{\text{uncond}} - V_{\text{reach}}
\end{equation}

Specifically, we prove a \textbf{Threshold Impossibility Theorem}: any compensation $C$ sufficient to close $\Delta V$ must necessarily induce a discontinuous categorical shift in the partner's identity, effectively bypassing the spread rather than closing it. This implies that financial transfers function as participation fees rather than utility substitutes in the presence of rigid spreads.

We further introduce a \textbf{dynamic discrete choice model} where commitment execution occurs when the market-to-book ratio $\theta$ crosses a time-decaying liquidity threshold $T$:
\begin{equation}
    \theta = \frac{V_{\text{reach}}}{V_{\text{uncond}}} \ge T(t) .
\end{equation}

This framework establishes a structural isomorphism between matching failures and \textbf{liquidity droughts}, and mathematically maps post-match regret to \textbf{execution slippage}. Numerical equilibrium analyses recover key empirical regularities, including the inefficiency of transfers in low-tier matches and the invariance of preference orderings under varying regional price norms. The results suggest that matching markets operate as \textbf{illiquid order-driven systems} where execution is constrained by inventory risk and reference-dependent spreads rather than price equilibrium.
}


\maketitle

\vspace{-2mm}
\begin{center}
\textbf{Code:} \url{https://github.com/Republic1024/Limit-Order-Matching-Microstructure}
\end{center}

\maketitle

\begin{figure*}[b!] 
    \centering
    \resizebox{1\textwidth}{!}{ 
    \begin{tikzpicture}[
        font=\sffamily\small,
        box/.style={draw, rounded corners, fill=white, align=center, drop shadow, line width=0.8pt},
        arrow/.style={->, >=Stealth, thick, color=gray!70},
        labeltext/.style={font=\bfseries\footnotesize, align=center},
        notetext/.style={font=\footnotesize, align=center, text width=2.2cm, color=black!80},
        tinytext/.style={font=\scriptsize, align=center, text width=1.8cm, color=black!70}
    ]
    \fill[blue!5] (-0.5, -4) rectangle (4.5, 4.5);
    \node[anchor=north west, text=blue!50!black, font=\bfseries] at (-0.3, 4.3) {A. Structural Gap ($\Delta V$)};
    
    \fill[green!5] (4.8, -4) rectangle (9.8, 4.5);
    \node[anchor=north west, text=green!50!black, font=\bfseries] at (5.0, 4.3) {B. Threshold Dynamics};
    
    \fill[orange!5] (10.1, -4) rectangle (15.1, 4.5);
    \node[anchor=north west, text=orange!50!black, font=\bfseries] at (10.3, 4.3) {C. Financial Isomorphism};

    
    \node[box, fill=red!10, minimum width=3cm, minimum height=0.8cm] (U1) at (2.0, 2.5) {\textbf{Ideal} ($V_{\text{uncond}}$)};
    \node[above, font=\scriptsize, text=red!50!black] at (U1.north) {Internal Expectation};
    
    \node[box, fill=blue!10, minimum width=3cm, minimum height=0.8cm] (R1) at (2.0, 0.0) {\textbf{Reality} ($V_{\text{reach}}$)};
    \node[below, font=\scriptsize, text=blue!50!black] at (R1.south) {Market Offer};
    
    \draw[<->, thick, red, dashed] (3.8, 2.5) -- (3.8, 0.0) node[midway, right, font=\footnotesize, align=left] {$\mathbf{\Delta V}$};
    \node[text=red!80!black, font=\scriptsize, align=center, text width=2cm] at (2.8, 1.25) {Structurally\\Unclosable};
    
    \node[draw, dashed, fill=yellow!20, minimum width=1.2cm, minimum height=0.6cm] (Comp) at (2.0, -2.0) {+$C$};
    \draw[->, thick] (Comp) -- (R1);
    \node[below, tinytext] at (Comp.south) {Compensation\\(Ineffective if $C < C^*$)};

    
    \draw[->, thick] (5.2, -1.5) -- (9.4, -1.5) node[right] {$t$}; 
    \draw[->, thick] (5.2, -1.5) -- (5.2, 3.0) node[above, text width=1cm, align=center] {Value}; 
    
    \draw[thick, red!80!black, domain=5.2:9.2] plot (\x, {2.5 - 0.3*(\x-5.2)^1.3}) node[above right, font=\scriptsize] {$T(t)$};
    \node[notetext, anchor=west] at (5.3, 0.5) {Age/Pressure\\lowers Threshold};
    
    \draw[thick, blue!80!black] (5.2, -0.5) 
        to[out=20, in=180] (7.0, 0.2) 
        to[out=0, in=200] (9.2, 1.8);
    \node[blue!80!black, font=\scriptsize] at (9.2, 2.0) {$\theta(t)$};
    
    \coordinate (Cross) at (8.35, 1.15);
    \fill[black] (Cross) circle (2pt);
    \draw[dashed] (Cross) -- (8.35, -1.5) node[below] {$t^*$};
    
    \node[coordinate, pin={[pin edge={black, thin}, align=center, font=\bfseries\scriptsize, pin distance=1.1cm]90:{Commit Event!\\($\theta \ge T$)}}] at (Cross) {};
    
    
    \node[box, fill=white, minimum width=3.8cm, align=center, inner sep=5pt] (OrderBook) at (12.6, 1.5) {
        \textbf{Limit Order Book} \\[0.2cm]
        \colorbox{red!15}{\makebox[3cm]{Ask: $V_{\text{uncond}}$}} \\ 
        \colorbox{white}{\makebox[3cm]{\textbf{Spread} $\approx \Delta V$}} \\ 
        \colorbox{green!15}{\makebox[3cm]{Bid: $V_{\text{reach}}$}}
    };
    
    \node[below=0.5cm of OrderBook, align=left, font=\footnotesize, text width=4.2cm] (Isomorphisms) {
        \textbf{Key Isomorphisms:}\\
        $\bullet$ \textbf{Regret} $\equiv$ Slippage\\
        $\bullet$ \textbf{Settling} $\equiv$ Market Order\\
        $\bullet$ \textbf{Rejection} $\equiv$ Wide Spread\\
        $\bullet$ \textbf{Shock} $\equiv$ Repricing
    };
    
    \draw[arrow, dashed] (9.8, 1.5) -- (10.7, 1.5);
    
    \end{tikzpicture}
    }
    \caption{\textbf{Graphical Abstract of the Unified Framework.} \textbf{(A)} The internal preference differential $\Delta V$ creates a structural gap between the Unconditional Max (Ideal) and Reachable Max (Reality) that monetary compensation ($C$) cannot structurally close. \textbf{(B)} Marriage decisions are governed by a state-machine dynamic where commitment occurs only when the reality-to-ideal ratio $\theta$ crosses the agent's willingness threshold $T$. \textbf{(C)} This system is structurally isomorphic to a financial Limit Order Book, where matching failures represent liquidity droughts due to wide bid-ask spreads.}
    \label{fig:graphical_abstract}
\end{figure*}
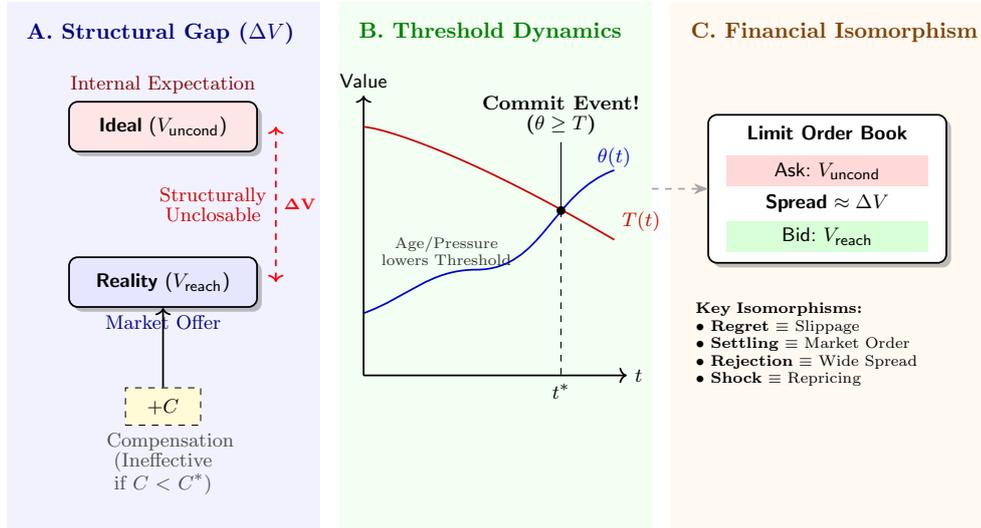

\section{Introduction}\label{sec1}

In the microstructure analysis of matching markets, a central question is whether continuous monetary transfers can effectively clear markets characterized by discrete, non-transferable utility (NTU) components. Classical economic models, following Becker \cite{becker1973}, postulate that \textit{price signals} (e.g., compensatory transfers or ``bride prices'') function as efficient clearing mechanisms: a lower-ranked agent can offer a sufficiently high transfer to offset a partner's utility deficit, thereby achieving equilibrium. Under this Transferable Utility (TU) assumption, liquidity is purely a function of price adjustment.

However, empirical evidence from high-friction matching markets contradicts this monotonicity assumption. High-tier agents frequently reject lower-tier offers despite substantial financial premiums, and compensation levels exhibit significant variance across regions with identical demographic structures \cite{anderson2007, chiappori2017}. These inefficiencies suggest that matching is not merely a price-mediated negotiation but a process constrained by \textbf{structural illiquidity} and \textbf{reference-dependent spreads}, analogous to wide bid--ask spreads in thinly traded asset markets \cite{demsetz1968cost, glosten1985bid}.

This paper proposes a \textbf{market microstructure framework} that formalizes mate selection as a \textbf{Limit Order Book (LOB) system} \cite{parlour2008limit}. We argue that the persistent failure of compensation to clear certain matches stems from an \textbf{internal preference differential}---a structural gap between an agent's internal reference price (what they believe is attainable) and the best available market bid. This aligns with behavioral findings on reference-dependent utility \cite{kahneman1979prospect, koszegi2006reference}.

Unlike price gaps in standard goods markets, this differential acts as a rigid \textbf{bid--ask spread} derived from ordinal rankings \cite{amihud1986asset}. Specifically, the agent's evaluation operates on a \textbf{two-sided Attribute Matrix}, defining a spread $\Delta V$:
\begin{equation}
\Delta V = V_{\text{uncond}} - V_{\text{reach}}.
\end{equation}
Here, $V_{\text{uncond}}$ represents the agent's internal \textbf{Ask Price} (Reservation Value based on counterfactuals), and $V_{\text{reach}}$ represents the market's best \textbf{Bid Price} (Reality). 

We formally demonstrate that this spread is \textbf{structurally immune to linear compensation}. Any financial transfer $C$ intended to close $\Delta V$ faces a \textbf{Bounded Utility Constraint}: unless $C$ is large enough to fundamentally reclassify the partner's categorical identity (a ``regime switch''), it functions merely as a price improvement within a rejected asset class, failing to trigger execution.

To capture the temporal dynamics of execution, we introduce a \textbf{dynamic discrete choice model} (a state-machine). Execution (marriage) occurs only when the market-to-book ratio $\theta$ crosses a time-decaying liquidity threshold $T$:
\begin{equation}
\frac{V_{\text{reach}}}{V_{\text{uncond}}} \ge T(t).
\end{equation}
This framework maps ``settling'' behavior to the \textbf{time decay of limit orders} due to inventory costs (e.g., age or social pressure) \cite{stoll1978supply}. Consequently, post-match dissatisfaction is mathematically equivalent to \textbf{execution slippage}---the realized negative spread between the target price and the execution price \cite{perold1988implementation}.

The contributions of this paper are threefold:
\begin{itemize}
    \item \textbf{Theoretical:} We establish a formal isomorphism between matching markets and financial Limit Order Books, demonstrating that ``psychological gaps'' function strictly as liquidity spreads.
    \item \textbf{Computational:} We implement an agent-based Attribute Matrix simulation to quantify the limits of compensatory transfers, extending classical matching theory with microstructure constraints \cite{todd1997}.
    \item \textbf{Empirical Consistency:} We show that the model recovers stylized facts of assortative mating---such as the ineffectiveness of transfers in low-tier matches and the persistence of slippage (regret)---viewing them as natural consequences of market microstructure rather than purely sociological phenomena.
\end{itemize}

By reframing the sociological problem of ``marrying down'' as a financial problem of \textbf{liquidity-constrained execution}, this work provides a unified computational theory for why money cannot always clear markets, and why structural spreads persist despite aggressive pricing signals.

\section{The Internal Differential Model}\label{sec2}

This section develops the mathematical core of our framework. We formalize (1) the agent's valuation structure, (2) the computation of the structural spread ($\Delta V$), and (3) the \textbf{Threshold Impossibility Theorem}, which proves that linear compensation cannot close spreads derived from ordinal categorization without inducing a regime switch.

\subsection{Preliminaries: Valuation Kernels and Attribute Matrices}\label{subsec2.1}

Let a female agent be denoted by $F$. Let the set of potential male partners in the market be
\begin{equation}
\mathcal{M} = \{M_1, M_2, \dots, M_n\}.
\end{equation}
Each candidate $M_i$ is evaluated by $F$ through a latent valuation kernel:
\begin{equation}
V_F(M_i) \in \mathbb{R}_+.
\end{equation}
This valuation represents a scalar compression of the candidate's row in the population \textbf{Attribute Matrix} (representing income, education, social capital).

We define two subsets within the agent's preference space:
\begin{enumerate}
    \item \textbf{Unconditional Option Set (The Ask Side):}
    \begin{equation}
    \mathcal{U}_F = \{ M_i : F \text{ believes agents like } M_i \text{ exist in the market} \}.
    \end{equation}
    \item \textbf{Reachable Option Set (The Bid Side):}
    \begin{equation}
    \mathcal{R}_F = \{ M_i : M_i \text{ is explicitly willing to choose } F \}.
    \end{equation}
\end{enumerate}
Let the maximal valuations in these sets be:
\begin{align}
V_{\text{uncond}} &= \max_{M \in \mathcal{U}_F} V_F(M), \\
V_{\text{reach}} &= \max_{M \in \mathcal{R}_F} V_F(M).
\end{align}
In microstructure terms, $V_{\text{uncond}}$ is the agent's \textbf{Internal Ask Price} (Reservation Price), and $V_{\text{reach}}$ is the \textbf{Best Available Bid}.

\subsection{The Structural Spread ($\Delta V$)}\label{subsec2.2}

We define the market friction as the \textbf{Internal Preference Differential}:
\begin{equation}\label{eq:delta_v_def}
\Delta V = V_{\text{uncond}} - V_{\text{reach}}.
\end{equation}
This $\Delta V$ functions as a rigid \textbf{Bid--Ask Spread}:
\begin{itemize}
    \item If $\Delta V = 0$: The market clears at the reservation price ($\text{Bid} \approx \text{Ask}$).
    \item If $\Delta V > 0$: There is a \textbf{Liquidity Deficit}. The agent perceives a loss of value ("trading down"), creating execution latency.
\end{itemize}

To clarify the mapping between this mathematical structure and social archetypes, we provide the conceptual mapping in Table \ref{tab:concept_mapping}.

\begin{table}[htbp]
\centering
\caption{\textbf{Microstructure Mapping of the Internal Preference Differential.}}
\label{tab:concept_mapping}
\renewcommand{\arraystretch}{1.5}
\begin{tabular}{p{0.15\textwidth} p{0.35\textwidth} p{0.4\textwidth}}
\hline
\textbf{Symbol} & \textbf{Microstructure Concept} & \textbf{Social Reality (Archetype)} \\ 
\hline
$V_{\text{uncond}}$ & \textbf{Internal Ask Price} & \textbf{``The Ideal / The CEO''} \newline The theoretical maximum value the agent believes they deserve based on market priors. \\ 
\hline
$V_{\text{reach}}$ & \textbf{Best Market Bid} & \textbf{``The Reality / The Suitor''} \newline The best actually available counterparty in the current order book. \\ 
\hline
$\Delta V$ & \textbf{Bid--Ask Spread} & \textbf{``The Gap / Structural Deficit''} \newline The source of friction. A positive spread implies the market cannot clear without slippage or price improvement. \\ 
\hline
$C$ & \textbf{Price Improvement} & \textbf{``Compensation / Bride Price''} \newline An additive transfer intended to facilitate execution despite a negative utility spread. \\
\hline
\end{tabular}
\end{table}

\begin{remark}
A positive $\Delta V$ implies a lack of immediate market clearing, reflecting frictions caused by information asymmetry or inventory risk \cite{glosten1985bid}. This necessitates either a spread-crossing mechanism (settling) or a liquidity timeout (threshold decay).
\end{remark}

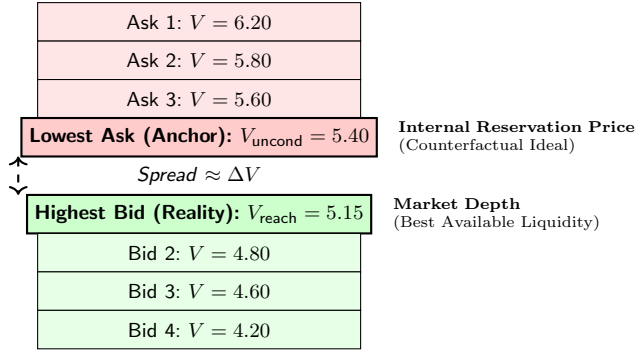
\begin{figure}[t]
    \centering
    \begin{tikzpicture}[scale=0.85, transform shape,
        box/.style={draw, rectangle, minimum width=5.0cm, minimum height=0.6cm, align=center},
        ask/.style={box, fill=red!10},
        bid/.style={box, fill=green!10},
        font=\sffamily\small
    ]
    \node[ask] (ask1) at (0, 2.4) {Ask 1: $V = 6.20$};
    \node[ask] (ask2) at (0, 1.8) {Ask 2: $V = 5.80$};
    \node[ask] (ask3) at (0, 1.2) {Ask 3: $V = 5.60$};
    \node[ask, fill=red!20, thick] (ask4) at (0, 0.6) {\textbf{Lowest Ask (Anchor):} $V_{\text{uncond}} = 5.40$};
    
    \node (gap) at (0, 0.0) {\textit{Spread} $\approx \Delta V$};
    \draw[<->, dashed, thick] (-2.8, 0.3) -- (-2.8, -0.3); 
    
    \node[bid, fill=green!20, thick] (bid1) at (0, -0.6) {\textbf{Highest Bid (Reality):} $V_{\text{reach}} = 5.15$};
    \node[bid] (bid2) at (0, -1.2) {Bid 2: $V = 4.80$};
    \node[bid] (bid3) at (0, -1.8) {Bid 3: $V = 4.60$};
    \node[bid] (bid4) at (0, -2.4) {Bid 4: $V = 4.20$};
    
    \node[right=0.2cm of ask4, text width=4.5cm, font=\footnotesize] 
        {\textbf{Internal Reservation Price} \\ (Counterfactual Ideal)};
        
    \node[right=0.2cm of bid1, text width=4.5cm, font=\footnotesize] 
        {\textbf{Market Depth} \\ (Best Available Liquidity)};
    \end{tikzpicture}
    \caption{\textbf{The Matching Market Order Book.} Visualizing the Internal Preference Differential $\Delta V$ as a rigid bid--ask spread. The market fails to clear because the highest bid ($V_{\text{reach}}$) does not cross the lowest ask ($V_{\text{uncond}}$).}
    \label{fig:order_book_structure}
\end{figure}

\subsection{Compensation as Price Improvement}\label{subsec2.3}

Let $C_i \ge 0$ denote a compensatory transfer (Price Improvement) offered by $M_i$. The effective utility becomes:
\begin{equation}
U_i = V_F(M_i) + f(C_i),
\end{equation}
where $f$ is a concave utility function representing diminishing returns ($f' > 0, f'' < 0$).

Crucially, the internal ask price $V_{\text{uncond}}$ is \textbf{invariant to external bids}. It is defined relative to the hypothetical "ideal" asset class. Thus, the effective spread after compensation is:
\begin{equation}
\Delta V(C) = V_{\text{uncond}} - (V_{\text{reach}} + f(C)).
\end{equation}
Since compensation is merely an additive term on the bid side, it does not alter the intrinsic valuation kernel $V_F(M_i)$. Therefore:
\begin{equation}
\Delta V(C) = \Delta V \quad \forall C < C^\star,
\end{equation}
where $C^\star$ is the \textbf{Regime Switching Threshold}.

\subsection{The Regime Switching Threshold $C^\star$}\label{subsec2.4}

We define the critical threshold:
\begin{equation}
C^\star = V_{\text{uncond}} - V_F(M_i).
\end{equation}
If $M_i$ offers $C \ge C^\star$, then $U_i \ge V_{\text{uncond}}$.
However, this condition implies a discontinuity. Compensation sufficient to close the gap effectively elevates the partner into the upper tier, rendering the compensation redundant as a "transfer" and redefining it as a signal of status elevation.

\subsection{The Threshold Impossibility Theorem}\label{subsec2.5}

We state the central theoretical result regarding market clearing via transfers.

\begin{theorem}[Threshold Impossibility]
For any partner $M$ with intrinsic value $V_F(M)$, if the compensation $C < C^\star$, then the structural spread $\Delta V$ remains invariant. The preference ordering is preserved:
\begin{equation}
U_i(C) < V_{\text{uncond}}.
\end{equation}
Only if $C \ge C^\star$ does the agent undergo a categorical regime switch. Thus, linear compensation cannot close a spread derived from ordinal categorization without inducing an identity collapse.
\end{theorem}

\subsubsection{Implications for Market Clearing}
\begin{itemize}
    \item \textbf{Invariance:} Financial transfers cannot modify the spread $\Delta V$ within the same asset class.
    \item \textbf{Non-Monotonicity:} Preferences do not respond continuously to price improvements; they exhibit step-function behavior at $C^\star$.
    \item \textbf{Liquidity Constraint:} The psychological gap acts as a hard liquidity constraint that money cannot "lubricate" unless the amount is transformative \cite{glosten1985bid}.
\end{itemize}

\subsection{Persistent Slippage Post-Execution}\label{subsec2.6}

Because $\Delta V$ is structural, even if execution occurs (due to threshold decay, see Sec 3), the spread remains encoded in the system:
\begin{align}
\lim_{t\to\infty} f(C(t)) &\to 0, \\
\Delta V(t) &= \Delta V.
\end{align}
This creates \textbf{Persistent Execution Slippage}: the realized utility is permanently below the target price. In the context of marriage, this slippage manifests as long-term dissatisfaction or regret, consistent with the behavior of traders executing at unfavorable prices to clear inventory \cite{stoll1978supply}.

\subsection{Zero-Spread Execution in High-Tier Matches}\label{subsec2.7}

For high-tier partners $M_{\text{top}}$:
\begin{equation}
V_F(M_{\text{top}}) \approx V_{\text{uncond}}.
\end{equation}
Thus, $\Delta V \approx 0$ and $C^\star \approx 0$.
This implies that high-tier matches are Zero-Spread Executions, requiring no price improvement ($C=0$). This explains the empirical regularity that the most stable matches often involve zero compensation, mirroring "at-the-market" trades in highly liquid assets.

\section{Dynamic Execution and Liquidity Thresholds}\label{sec3}

While Section 2 formalized the static structure of the spread $\Delta V$, real matching decisions unfold dynamically over time \cite{todd1997}. Agents do not make a single one-shot decision; instead, execution thresholds evolve as the order book updates and inventory costs (e.g., age) accumulate \cite{mortensen1986}. In this section, we formalize the matching process as a \textbf{dynamic discrete choice model}.

\subsection{States, Transitions, and the Search Process}\label{subsec3.1}

Let the agent $F$ exist in one of three macro-states:
\begin{equation}
S \in \{ S_{\text{search}},\; S_{\text{evaluate}},\; S_{\text{execute}} \}.
\end{equation}

\subsubsection{State definitions}
\begin{itemize}
    \item \textbf{$S_{\text{search}}$ (Open Order)}: The agent is active in the market, scanning the order book. The candidate set updates stochastically.
    \item \textbf{$S_{\text{evaluate}}$ (Price Discovery)}: The agent narrows the candidate set and computes the market-to-book ratio ($\theta$). This is equivalent to checking if a limit order is marketable \cite{glosten1985bid}.
    \item \textbf{$S_{\text{execute}}$ (Commitment)}: The trade is executed (marriage). This is an absorbing state unless significant external shocks trigger a "regime switch" (divorce).
\end{itemize}
Transitions are governed by a time-varying \textbf{Liquidity Threshold} $T(t)$.

\subsection{The Liquidity Threshold $T(t)$}\label{subsec3.2}

We define the decision threshold $T \in (0,1]$, representing the agent's \textbf{urgency to execute}. In microstructure terms, this reflects the inventory holding cost \cite{stoll1978supply}. Factors influencing $T$ include:
\begin{itemize}
    \item \textbf{Inventory Age} (increases urgency $\to$ lowers $T$).
    \item \textbf{Market Volatility} (uncertainty increases $T$).
    \item \textbf{Peer Pressure} (social signals act as inventory cost shocks).
\end{itemize}
$T$ evolves according to a decay function:
\begin{equation}
T_{t+1} = g(T_t, \; \text{inventory cost}, \; \text{volatility}),
\end{equation}
where $\frac{\partial g}{\partial t} < 0$, implying that limit orders become more aggressive (lower ask) over time to ensure execution.

\subsection{The Execution Condition}\label{subsec3.3}

At evaluation time $t$, the agent computes the Market-to-Book Ratio:
\begin{equation}\label{eq:theta_def}
\theta_t = \frac{V_{\text{reach}}(t)}{V_{\text{uncond}}(t)}
\end{equation}
This ratio $\theta_t$ measures how close the best market bid is to the internal reservation price.

\subsubsection{Execution occurs iff:}
\begin{equation}
\theta_t \ge T_t.
\end{equation}
\textbf{Interpretation}: If the market bid is "close enough" to the internal ask (crossing the liquidity threshold), the trade executes. If the spread is too wide relative to urgency, the order remains open.

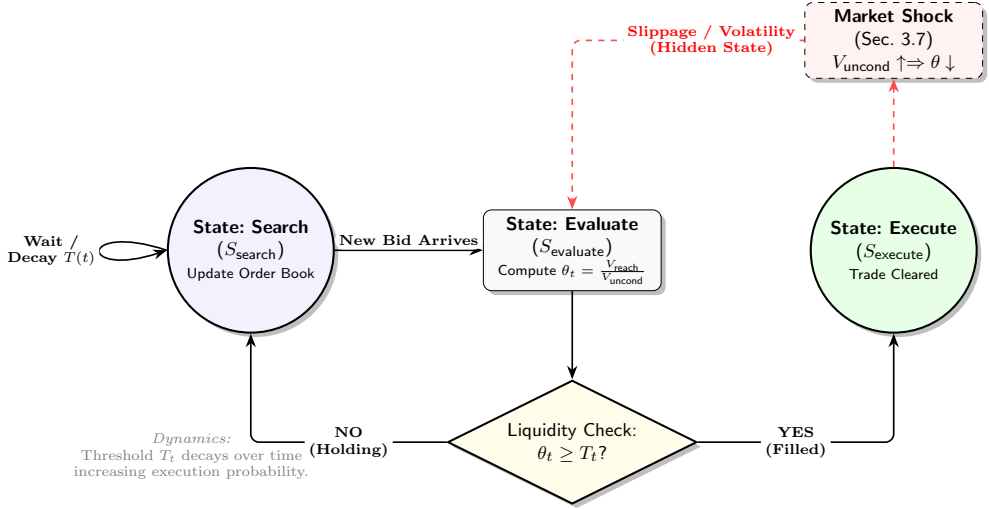
\begin{figure*}[t!]
    \centering
    \resizebox{1.0\textwidth}{!}{ 
    \begin{tikzpicture}[
        node distance=2cm and 2cm,
        font=\sffamily\small,
        state/.style={draw, circle, minimum size=2.8cm, align=center, fill=blue!5, line width=1pt, drop shadow},
        decision/.style={draw, diamond, aspect=2, minimum width=3cm, minimum height=1.5cm, align=center, fill=yellow!10, line width=1pt, drop shadow},
        process/.style={draw, rectangle, rounded corners, minimum width=3cm, minimum height=1cm, align=center, fill=gray!5, drop shadow},
        arrow/.style={->, >=Stealth, thick, rounded corners},
        labeltext/.style={font=\footnotesize\bfseries, align=center, fill=white, inner sep=2pt}
    ]

    
    \node[state] (Search) at (0, 0) {\textbf{State: Search}\\($S_{\text{search}}$)\\ \footnotesize Update Order Book};
    
    \node[process, right=2.5cm of Search] (Evaluate) {\textbf{State: Evaluate}\\($S_{\text{evaluate}}$)\\ \footnotesize Compute $\theta_t = \frac{V_{\text{reach}}}{V_{\text{uncond}}}$};
    
    \node[decision, below=1.5cm of Evaluate] (Decide) {Liquidity Check:\\$\theta_t \ge T_t$?};
    
    \node[state, fill=green!10, right=2.5cm of Evaluate] (Commit) {\textbf{State: Execute}\\($S_{\text{execute}}$)\\ \footnotesize Trade Cleared};
    
    \node[process, dashed, fill=red!5, above=1.5cm of Commit] (Shock) {\textbf{Market Shock}\\(Sec. 3.7)\\$V_{\text{uncond}} \uparrow \Rightarrow \theta \downarrow$};


    \draw[arrow] (Search) -- node[labeltext, above] {New Bid Arrives} (Evaluate);

    \draw[arrow] (Evaluate) -- (Decide);

    \draw[arrow] (Decide.east) -| node[labeltext, near start] {YES\\(Filled)} (Commit.south);

    \draw[arrow] (Decide.west) -| node[labeltext, near start] {NO\\(Holding)} (Search.south);

    \draw[arrow, dashed, red!70] (Commit.north) -- (Shock.south);
    \draw[arrow, dashed, red!70] (Shock.west) -| node[labeltext, pos=0.2, text=red] {Slippage / Volatility\\(Hidden State)} (Evaluate.north);

    \draw[arrow] (Search.west) .. controls +(-1.5, 0.5) and +(-1.5, -0.5) .. node[labeltext, left] {Wait /\\Decay $T(t)$} (Search.west);

    
    \node[text width=4cm, align=center, font=\footnotesize, color=gray] at (-1, -3.5) {
        \textit{Dynamics:}\\
        Threshold $T_t$ decays over time\\
        increasing execution probability.
    };
    \end{tikzpicture}
    } 
    \caption{\textbf{Dynamic Execution Process.} The diagram illustrates the agent's decision flow. Execution occurs when the market-to-book ratio $\theta_t$ crosses the time-decaying liquidity threshold $T_t$. External shocks (e.g., peer comparison) can trigger post-execution volatility by repricing the internal ask.}
    \label{fig:state_machine}
\end{figure*}

\subsection{Execution under Negative Spread (Threshold Decay)}\label{subsec3.4}

A counterintuitive result is that execution often occurs even when the spread is positive ($\Delta V > 0$).
\begin{equation}
\Delta V = V_{\text{uncond}} - V_{\text{reach}} > 0 \quad \text{but} \quad \theta_t > T_t.
\end{equation}
\textbf{Mechanism}: The agent executes not because the spread closed, but because the liquidity threshold $T_t$ decayed sufficiently to accept the current bid.
This explains the phenomenon of "settling" not as a preference shift, but as a rational response to \textbf{liquidity constraints}. Compensation ($C$) is not needed to close $\Delta V$; rather, time decay lowers the barrier to execution.

\subsection{Immediate Execution in High-Tier Matches}\label{subsec3.5}

If the agent encounters a top-tier bid $M_{\text{top}}$, then $V_{\text{reach}} \approx V_{\text{uncond}}$, implying $\theta_t \approx 1$.
Since $T_t \le 1$ by definition, the condition $\theta_t > T_t$ is satisfied instantaneously.
Therefore: $S_{\text{search}} \rightarrow S_{\text{evaluate}} \rightarrow S_{\text{execute}}$ occurs within a single time step. This models the \textbf{immediate fill} (marketable limit order) observed in high-compatibility matches \cite{parlour2008limit}.

\subsection{Persistent Slippage Post-Execution}\label{subsec3.6}

Post-execution:
\begin{itemize}
    \item Compensation utility $f(C)$ dissipates (one-time transfer).
    \item The structural spread $\Delta V$ remains encoded.
    \item The internal ask $V_{\text{uncond}}$ persists as a reference point.
\end{itemize}
Thus, the ratio $\theta_t < 1$ persists. This implies that the trade was executed at a price below the initial target, resulting in \textbf{permanent execution slippage}. In social terms, this manifests as "post-marital dissatisfaction" or regret \cite{perold1988implementation}.

\subsection{External Shocks and Repricing (Volatility)}\label{subsec3.7}

Post-execution, external information arrival (e.g., observing a peer's superior match) can act as a market shock, shifting the internal ask price upward:
\begin{equation}
V_{\text{uncond}}(t+1) > V_{\text{uncond}}(t).
\end{equation}
Assuming the bid $V_{\text{reach}}$ (current partner) is fixed (illiquid), the ratio $\theta$ drops sharply. If $\theta_{t+1} < T_{t+1}$, the agent enters a state of \textbf{hidden volatility} or dissatisfaction. This models the dynamic instability of matches exposed to social comparison shocks.

\subsection{Summary of the Dynamic Model}\label{subsec3.8}
The core logic is:
\begin{enumerate}
    \item \textbf{Spread Persistence:} The internal spread $\Delta V$ is structural and immune to linear compensation unless $C \ge C^\star$.
    \item \textbf{Threshold Execution:} Decisions are driven by the liquidity threshold $T(t)$, not just price maximization.
    \item \textbf{Time Decay:} As inventory costs rise (age), $T(t)$ decays, facilitating execution even with positive spreads ("settling").
    \item \textbf{Immediate Fills:} High-quality bids ($\theta \approx 1$) trigger instant execution.
    \item \textbf{Slippage:} Execution below the ask ($\theta < 1$) results in persistent structural regret.
    \item \textbf{Volatility Shocks:} External information reprices $V_{\text{uncond}}$, destabilizing the equilibrium.
\end{enumerate}

\section{Numerical Analysis of Execution Dynamics}\label{sec4}

To evaluate the implications of the Internal Differential Model and the dynamic execution rules, we conduct a series of numerical experiments. These simulations function as an agent-based stress test of the model, asking a core microstructure question: \textit{Under what conditions does price improvement ($C$) trigger execution, and under what conditions is the spread structurally invariant?}

\subsection{Model Calibration and Setup}\label{subsec4.1}

We simulate a heterogeneous population of agents where preferences are encoded in a vectorized Attribute Matrix.

\subsubsection{Distributional Assumptions}
Unlike standard models assuming a Gaussian distribution of utility, we model the intrinsic value $V_i$ using a Beta Distribution (mimicking the Gini Conical Structure described in Appendix F). This reflects the geometric scarcity of high-tier liquidity. Each agent $j$ computes their own idiosyncratic spread $\Delta V^{(j)} = V_{\text{uncond}}^{(j)} - V_{\text{reach}}^{(j)}$.

\subsubsection{Bounded Price Improvement Rule}
An agent can offer compensation $C$ (Price Improvement). The effective utility is $\tilde{V}_i = V_i + h(C)$, where the utility function is subject to a Bounded Constraint:
\begin{equation}
h(C) = \min(C \cdot \epsilon, C^\star_{\text{cap}}).
\end{equation}
This "clipping function" encodes the Regime Switching Threshold: financial transfers can improve the bid price within a class but cannot bridge structural categorical gaps unless the amount is transformative.

\subsection{Experiment 1: Ineffectiveness of Liquidity Injection in Deep Out-of-the-Money Matches}\label{subsec4.2}

\textbf{Hypothesis:} Does aggressive price improvement clear a market with a deep structural spread?

\textbf{Setup:}
\begin{itemize}
    \item Internal Ask: $V_{\text{uncond}}=95$
    \item Deep OTM Bid: $V_{\text{reach}} = 60$
    \item Price Improvement: $C = 500$k (Aggressive Liquidity Injection)
    \item Utility Cap: $C^\star_{\text{cap}} = 20$ utility points
\end{itemize}

\textbf{Equilibrium Snapshot:}
\begin{quote}
\small
\textbf{Status:} Execution Failed (Order Rejected) \\
Effective Bid: $U = 60 + \min(500\times 0.05, 20) = 60 + 20 = 80$ \\
Market-to-Book Ratio: $\theta = 80/95 = 0.84$ \\
\textbf{Result:} With a standard liquidity threshold $T=0.90$, execution fails despite the 500k offer. \\
\textbf{Realized Spread:} $95 - 80 = 15.0$ (Structural Liquidity Deficit).
\end{quote}

\textbf{Conclusion:} This validates the Threshold Impossibility Theorem: The spread $\Delta V$ acts as a hard constraint. The failure to clear is due to the "clipping" of financial utility, confirming that money cannot substitute for asset quality beyond a bounded range.

\subsection{Experiment 2: Execution via Threshold Decay ("Settling")}\label{subsec4.3}

\textbf{Hypothesis:} How does time-decay of the limit order ($T$) facilitate execution despite a positive spread?

\textbf{Setup:}
\begin{itemize}
    \item $V_{\text{uncond}} = 90$
    \item $V_{\text{reach}} = 70$
    \item $\Delta V = 20$ (Uncompensated Spread)
    \item Market-to-Book Ratio: $\theta = 70/90 \approx 0.78$
\end{itemize}

We introduce a liquidity threshold $T(t)$ that decays with inventory age (Table~\ref{tab:age_threshold}).

\begin{table}[h]
\centering
\caption{Time Decay of Liquidity Threshold $T(t)$ due to Inventory Cost.}
\label{tab:age_threshold}
\begin{tabular}{lccccc}
\toprule
Time (proxy for Age)      & $t_1$   & $t_2$   & $t_3$   & $t_4$   & $t_5$   \\
\midrule
$T$ (Threshold)  & 0.95 & 0.88 & 0.80 & \textbf{0.75} & 0.70 \\
\bottomrule
\end{tabular}
\end{table}

\textbf{Equilibrium Snapshot:}
\begin{quote}
\small
Time $t_1$: $\theta (0.78) < T (0.95) \Rightarrow$ Status: \textbf{HOLD} \\
Time $t_3$: $\theta (0.78) < T (0.80) \Rightarrow$ Status: \textbf{HOLD} \\
Time $t_4$: $\theta (0.78) > T (0.75) \Rightarrow$ Status: \textbf{EXECUTE} \\
\end{quote}

\textbf{Conclusion:} Execution at $t_4$ is driven by threshold decay, not price improvement. This confirms that "settling" is a rational response to inventory holding costs in an illiquid market.

\subsection{Experiment 3: Immediate Fills with High-Quality Liquidity}\label{subsec4.4}

\textbf{Setup:}
\begin{itemize}
    \item $V_{\text{uncond}} = 90$
    \item Incoming Bid: $V=94$ (High Tier), $C=0$
\end{itemize}

\textbf{Equilibrium Snapshot:}
\begin{quote}
\small
$\theta = 94/90 = 1.04$ \\
Condition: $\theta > T$ is satisfied for \textbf{ANY} rational $T \le 1$. \\
Result: \textbf{Immediate Fill} (Marketable Limit Order).
\end{quote}

\textbf{Conclusion:} High-quality assets bypass the threshold decay process entirely, executing instantaneously as marketable orders.

\subsection{Experiment 4: Invariance to Regional Pricing Norms}\label{subsec4.5}

\textbf{Hypothesis:} Do regional differences in base compensation (price norms) alter the ordinal ranking of the order book?

\textbf{Setup:} Two markets with identical asset pools but different norms:
\begin{itemize}
    \item Market A (High-Cost): $C_{\text{base}} = 200$k
    \item Market B (Low-Cost): $C_{\text{base}} = 30$k
\end{itemize}

\textbf{Equilibrium Snapshot:}
\begin{quote}
\small
Market A Best Execution: Counterparty A ($V=85$) \\
Market B Best Execution: Counterparty A ($V=85$)
\end{quote}

\textbf{Interpretation:} Although the absolute transfer amounts differ significantly, the ordinal ranking of the order book remains invariant. Compensation acts as a market intercept (participation fee), not a slope that reorders preferences.

\textbf{Conclusion:} Preference structures are robust to nominal price levels; relative spreads determine execution.

\subsection{Experiment 5: Execution Slippage (Post-Match Regret)}\label{subsec4.6}

\textbf{Setup:} Post-execution, the bid $V_{\text{reach}}$ is fixed (illiquid). We simulate an external informational shock (e.g., peer comparison) that reprices the internal ask.

\textbf{Equilibrium Snapshot:}
\begin{quote}
\small
\textbf{State 0 (Executed):} Bid $=75$, Ask $=90$ ($\theta=0.83 > T=0.80$). Stable. \\
\textbf{Shock:} Peer executes at $V=99$. \\
\textbf{Repricing:} New $V_{\text{uncond}} \leftarrow 99$. \\
\textbf{State 1 (Slippage):} New $\theta = 75/99 = 0.76$. \\
Condition: $0.76 < T(0.80) \Rightarrow$ \textbf{Regime Switch (Volatility/Regret)}.
\end{quote}

\textbf{Conclusion:} Regret is mathematically equivalent to Execution Slippage. It occurs when the realized Market-to-Book ratio $\theta$ drops below the historical liquidity threshold $T$ due to the upward repricing of the internal ask.

\section{Empirical Predictions}\label{sec5}
The Internal Differential Model and the Dynamic Execution Framework yield a set of \textbf{testable, quantitative microstructure predictions} derived from the structural rigidity of the spread.

\subsection{Prediction Set A: Limits of Price Improvement}
\begin{itemize}
    \item \textbf{A1 (Ordinal Invariance):} Continuous monetary transfers (price improvement) will not alter the ordinal ranking of the order book. If Bid $A <$ Bid $B$ in intrinsic value, then $A+C$ will not execute over $B$ unless $C$ triggers a regime switch ($C \ge C^\star$).
    \item \textbf{A2 (Non-Linearity):} The elasticity of execution probability to compensation is highly non-linear, exhibiting a step-function behavior at the categorical threshold $C^\star$.
    \item \textbf{A3 (Persistent Spread):} Post-execution surveys will reveal a persistent "psychological gap" ($\Delta V > 0$) in compensated matches, quantifying the structural inefficiency of the clearing mechanism.
\end{itemize}

\subsection{Prediction Set B: Regional Norms as Base Fees}
\begin{itemize}
    \item \textbf{B1 (Ranking Invariance):} Agent ranking preferences will remain invariant across regions with different base compensation norms (e.g., high-cost vs. low-cost marriage markets).
    \item \textbf{B2 (Intercept vs. Slope):} Regional norms affect the \textit{intercept} of the transaction cost but not the \textit{slope} of the preference function.
\end{itemize}

\subsection{Prediction Set C: Inventory Costs and Threshold Decay}
\begin{itemize}
    \item \textbf{C1 (Time Decay):} "Age" (proxy for inventory holding time) will be a significant negative predictor of the liquidity threshold $T$, independent of asset quality.
    \item \textbf{C2 (Inflection Point):} The execution probability function will exhibit a "threshold inflection point" (e.g., age 30-35) where the decay of $T$ accelerates to clear inventory.
    \item \textbf{C3 (Spread Acceptance):} Agents with higher inventory costs will accept wider realized spreads ($\Delta V$), confirming that "settling" is a rational liquidity response.
\end{itemize}

\subsection{Prediction Set D: Marketable Limit Orders (High-Tier)}
\begin{itemize}
    \item \textbf{D1 (Immediate Fills):} If a bid approaches the internal ask ($V_{\text{bid}} \approx V_{\text{ask}}$), execution latency will drop to near-zero ("instant commitment"), regardless of compensation.
    \item \textbf{D2 (Volatility Shocks):} Exposure to high-value third-party signals (e.g., peers executing better trades) will destabilize existing contracts by inflating the internal ask $V_{\text{uncond}}$.
\end{itemize}

\subsection{Prediction Set E: Execution Slippage (Regret)}
\begin{itemize}
    \item \textbf{E1 (Slippage Probability):} The probability of post-execution regret is directly proportional to the pre-execution spread $\Delta V$.
    \item \textbf{E2 (Repricing Risk):} Informational shocks that raise the reference price $V_{\text{uncond}}$ will cause the realized market-to-book ratio $\theta$ to plummet, potentially triggering a "stop-loss" event (divorce).
\end{itemize}

\section{Discussion}\label{sec6}
This paper introduces a unified microstructure framework for understanding matching markets through \textbf{internal preference differentials}, \textbf{dynamic execution thresholds}, and \textbf{bounded compensation mechanisms}. By formalizing individual choice as a function of internal counterfactuals (Ask Price) rather than external market prices, this model reveals the deeper structural constraints of non-clearing markets.

\subsection{Insight 1: Money Cannot Close Structural Spreads}
Traditional TU models assume that utility deficits can be arbitraged away by transfers. Our model demonstrates that compensation has zero marginal effect on the ordinal ranking of assets:
\begin{equation}
    A < B \implies A + C \not\succ B \quad (\text{if } C < C^\star)
\end{equation}
unless the transfer reaches a Regime Switching Threshold $C^\star$. This resolves the empirical puzzle of why high premiums fail to clear low-tier inventory. The answer lies in the discrete nature of the asset class: compensation modifies the transaction price, but the asset quality is governed by the counterfactual baseline. Thus, transfers act as liquidity facilitators, not utility substitutes.

\subsection{Insight 2: Execution is Driven by Threshold Decay}
We reframe the concept of "settling" not as a change in preferences, but as a rational adjustment of the execution threshold $T(t)$ in response to inventory costs.
\begin{equation}
\theta = \frac{V_{\text{bid}}}{V_{\text{ask}}} \ge T(t)
\end{equation}
This dynamic explains non-linear phenomena such as the sudden acceleration of matching rates at certain age cohorts and the "panic selling" behavior observed when liquidity dries up.

\subsection{Insight 3: Structural Isomorphism with Limit Order Books}
The most significant theoretical contribution is the identification of a structural isomorphism between social matching and financial microstructure (see detailed mapping in Appendix D).
\begin{itemize}
    \item Spread $\Delta V$ $\equiv$ Bid--Ask Spread: Represents the cost of immediate execution vs. waiting.
    \item Regret $\equiv$ Execution Slippage: The mathematical difference between the target price and the realized price.
    \item Social Shocks $\equiv$ Information Arrival: Updates to the internal order book that reprice the Ask side.
\end{itemize}
This implies that matching markets are subject to the same laws of liquidity, volatility, and market depth as financial markets.

\subsection{Insight 4: Meso-Scale Constraints (Buckets and Cones)}
Our framework integrates micro-decisions with meso-scale constraints:
\begin{itemize}
    \item Behavioral Heterogeneity (Appendix E): Agents do not optimize over a continuous distribution but rely on heuristic "Prior Liquidity Buckets." This discretization explains rigid pricing regimes.
    \item Structural Scarcity (Appendix F): We model the socio-economic hierarchy as a "Gini Conical Structure." This geometric model proves that the available volume of reachable partners shrinks super-linearly as status rises, creating a mathematically necessary Liquidity Drought at the top of the pyramid.
\end{itemize}

\subsection{Theoretical Contributions}
This work provides:
\begin{enumerate}
    \item \textbf{A Unified Microstructure Model:} Bridging matching theory, decision theory, and market microstructure.
    \item \textbf{Resolution of the Compensation Paradox:} Explaining why transfers cannot clear structural spreads without identity transformation.
    \item \textbf{Quantification of Slippage:} Providing a mathematical definition for post-match dissatisfaction based on spread analysis.
    \item \textbf{A General Theory of Illiquid Matching:} Applicable beyond marriage to labor markets, university admissions, and OTC asset trading.
\end{enumerate}

\subsection{Limitations and Extensions}
Future work can refine several areas:
\begin{itemize}
    \item \textbf{L1. Volatility Estimation:} Modeling the variance of $V_{\text{uncond}}$ as a stochastic volatility process.
    \item \textbf{L2. Network Effects:} Explicitly modeling the propagation of informational shocks through a social network graph.
    \item \textbf{L3. Multi-Period Games:} Extending the single-execution model to a repeated game with renegotiation (divorce/remarriage options).
\end{itemize}


\section{Conclusion}\label{sec7}
By formalizing mate selection as a \textbf{Limit Order Book (LOB) system} rather than a cooperative game or a transferable utility market, this paper provides a microstructure explanation for the limits of compensatory transfers. We demonstrate that the internal preference differential $\Delta V$ acts as a structural \textbf{bid--ask spread} that linear price improvement (compensation) cannot bridge without triggering a categorical regime switch. Matches in this system are governed not by market-clearing prices, but by the dynamics of \textbf{threshold decay} and \textbf{inventory management}.

The model reveals that post-match dissatisfaction is structurally isomorphic to \textbf{execution slippage} in illiquid markets: the realized spread between the internal Ask and the executed Bid persists due to the invariance of the reference point. \textbf{Ultimately, we show that execution probabilities are bounded by dual structural constraints: the geometric \textit{Liquidity Drought} imposed by the ``Gini Cone'' (Appendix F) and the rigid \textit{Pricing Regimes} derived from ``Prior Liquidity Buckets'' (Appendix E).}

This unified framework bridges market microstructure and social matching theory, offering a rigorous quantitative basis for analyzing \textbf{non-clearing markets} characterized by structural spreads, inventory risk, and reference-dependent execution.

\backmatter

\bmhead{Supplementary Information}
Detailed proofs, attribute matrix configurations, and simulation source code are provided in Appendices A, B, and C.

\bmhead{Acknowledgements}
The author thanks the anonymous reviewers for their insights regarding the structural isomorphism between matching markets and limit order book dynamics.

\section*{Declarations}
\begin{itemize}
\item \textbf{Funding:} Not applicable.
\item \textbf{Conflict of interest:} The authors declare no competing interests.
\item \textbf{Data availability:} Theoretical and computational study; no empirical data generated.
\item \textbf{Code availability:} The Python simulation code used for the numerical equilibrium analysis is available in Appendix C.
\end{itemize}


\begin{appendices}

\section{Example Latent Preference State Matrix (LPSM)}\label{secA1}
This appendix provides a numerical example of the agent's internal \textbf{Attribute Matrix} (LPSM), demonstrating the computation of the internal ask ($V_{\text{uncond}}$), the best bid ($V_{\text{reach}}$), the spread ($\Delta V$), and the market-to-book ratio ($\theta$).

\subsection{Scenario Setup: Order Book Initialization}
Consider an agent $F$ managing a portfolio of potential counterparties. The preference space is discretized into asset classes based on valuation kernels (see Table \ref{tab:appendix_A1}).

\begin{table}[h]
\caption{Agent F's Latent Preference State Matrix (LPSM).}\label{tab:appendix_A1}
\begin{tabular}{@{}lllll@{}}
\toprule
ID & Asset Class & Availability & Liquidity Status & Valuation $V_F(M_i)$ \\
\midrule
H  & Prime Asset (e.g., CEO) & Hypothetical & Illiquid (Ask Side) & 95 \\
B  & High-Grade (Engineer) & Exists & Lock-up (Attached) & 88 \\
C  & Mid-Cap (Developer) & Exists & Liquid (Bid Side) & 78 \\
D  & Stable Value (Civil Servant) & Exists & Liquid (Bid Side) & 72 \\
E  & Distressed Asset & Exists & Liquid (Bid Side) & 60 \\
\botrule
\end{tabular}
\end{table}

We derive the structural parameters from this matrix:
\begin{itemize}
    \item \textbf{Internal Ask Price (Reservation):} $V_{\text{uncond}} = \max(H) = 95$
    \item \textbf{Best Market Bid (Execution):} $V_{\text{reach}} = \max(C, D, E) = 78$
    \item \textbf{Structural Bid-Ask Spread:} $\Delta V = 95 - 78 = 17$
    \item \textbf{Market-to-Book Ratio:} $\theta = 78 / 95 \approx 0.82$
\end{itemize}

\subsection{Compensation as Price Improvement}
Assume the following Price Improvement orders ($C$) are submitted:
\begin{itemize}
    \item Bid C: $C_C = 200$k (Moderate)
    \item Bid D: $C_D = 300$k (Aggressive)
    \item Bid E: $C_E = 0$ (At Market)
\end{itemize}
Using a linear utility function with a bounded constraint ($U_i = V_F(M_i) + \alpha \cdot C_i$, $\alpha = 0.05$):
\begin{itemize}
    \item Effective Bid C: $78 + 4 = 82$
    \item Effective Bid D: $72 + 6 = 78$
    \item Effective Bid E: $60 + 0 = 60$
\end{itemize}
\textbf{Result:} The agent selects Bid C as the optimal execution target.
\textbf{Microstructure Implication:} Note that $V_{\text{uncond}} = 95$ remains invariant. The effective spread narrows to $95 - 82 = 13$, but does not close. The execution slippage remains positive.

\subsection{Execution Logic under Threshold Decay}
Assume the agent's current Liquidity Threshold is $T(t) = 0.80$.
Since the realized ratio $\theta \approx 0.82 > 0.80$, the execution condition is met.
The order is filled (Commitment State).
\begin{itemize}
    \item \textbf{Executed Price:} 82 (Valuation + Transfer)
    \item \textbf{Inventory Cost:} High (driving $T$ down to 0.80)
    \item \textbf{Realized Slippage:} 13 units (Persistent)
\end{itemize}
This formalizes "settling" as a rational execution decision under inventory constraints, rather than a preference reversal.

\section{Counterparty Valuation and Dual-Sided Matching}\label{secA2}
This appendix presents the counterparty's (Male agent $M$) side of the order book. The market is two-sided; execution requires a double coincidence of wants satisfying dual liquidity constraints.

\subsection{Counterparty Preference Structure}
We define the male agent's sets:
\begin{itemize}
    \item $\mathcal{U}_M$: Unconditional Set (Ask Side)
    \item $\mathcal{R}_M$: Reachable Set (Bid Side)
\end{itemize}
This yields the counterparty's own spread metrics:
\begin{itemize}
    \item Internal Ask: $V_{\text{uncond}}^{(M)} = \max_{F \in \mathcal{U}_M} V_M(F)$
    \item Best Bid: $V_{\text{reach}}^{(M)} = \max_{F \in \mathcal{R}_M} V_M(F)$
    \item Spread: $\Delta V_M = V_{\text{uncond}}^{(M)} - V_{\text{reach}}^{(M)}$
\end{itemize}

\subsection{Circuit Breaker: The Compensation Ceiling $C_{\max}$}
When submitting a Price Improvement order to agent $F^*$, the male $M$ computes his maximum willingness to pay, $C_{\max}(F^*)$. This functions as a Circuit Breaker or Stop-Loss Limit, incorporating wealth constraints and ROI analysis.
\begin{equation}
C_{\max}(F^*) = \arg\max_{C} \; \Big\{  U_M(F^*, C) - \text{TransactionCost}(C) \Big\},
\end{equation}
When agent $F$ posts an Ask Price implying a required compensation $C_{\text{ask}}$:
\begin{itemize}
    \item If $C_{\text{ask}} \le C_{\max}(F^*)$: The order is valid.
    \item If $C_{\text{ask}} > C_{\max}(F^*)$: The order is \textbf{Rejected}. A regime switch occurs ("Walk Away"), and the match fails.
\end{itemize}
This constraint explains why aggressive ask prices often lead to liquidity dry-ups rather than higher clearing prices.

\subsection{Coupled Execution: The Dual-LOB Interaction}
A match $(F, M)$ executes if and only if the following Triple Coincidence Constraint is satisfied:
\begin{enumerate}
    \item \textbf{F's Liquidity Check:} $\theta_F = \frac{V_{\text{reach}}^{(F)}}{V_{\text{uncond}}^{(F)}} \ge T_F$
    \item \textbf{M's Liquidity Check:} $\theta_M = \frac{V_{\text{reach}}^{(M)}}{V_{\text{uncond}}^{(M)}} \ge T_M$
    \item \textbf{Capital Constraint:} $C_{\text{required}}(F, M) \le C_{\max}(M)$
\end{enumerate}
Failure of any condition results in a failed trade (No Match). This model captures the asymmetry where one side is willing to execute (liquidity provided) but the other side holds out (liquidity withheld), or where capital constraints prevent clearing despite mutual interest.

\section{Python-based Simulations: The Gini-Cone Logic}\label{sec:appendix_c}

This appendix details the computational implementation of the \textit{Internal Differential Model}. We model the agent population using a vectorized approach, where the \textbf{Attribute Matrix} is implemented as a Pandas DataFrame for efficient column-wise operations.

Unlike standard models that assume a Gaussian distribution of partner value, this simulation employs a \textbf{Beta Distribution} (mimicking the Pareto/Gini structure) to reflect the geometric scarcity of high-tier partners. It explicitly calculates the ``Slippage'' (Regret) and enforces the ``Identity Collapse Threshold'' ($C_{max}$) derived in Theorem 1.

\begin{figure}[htbp]
    \centering
    \includegraphics[width=\linewidth]{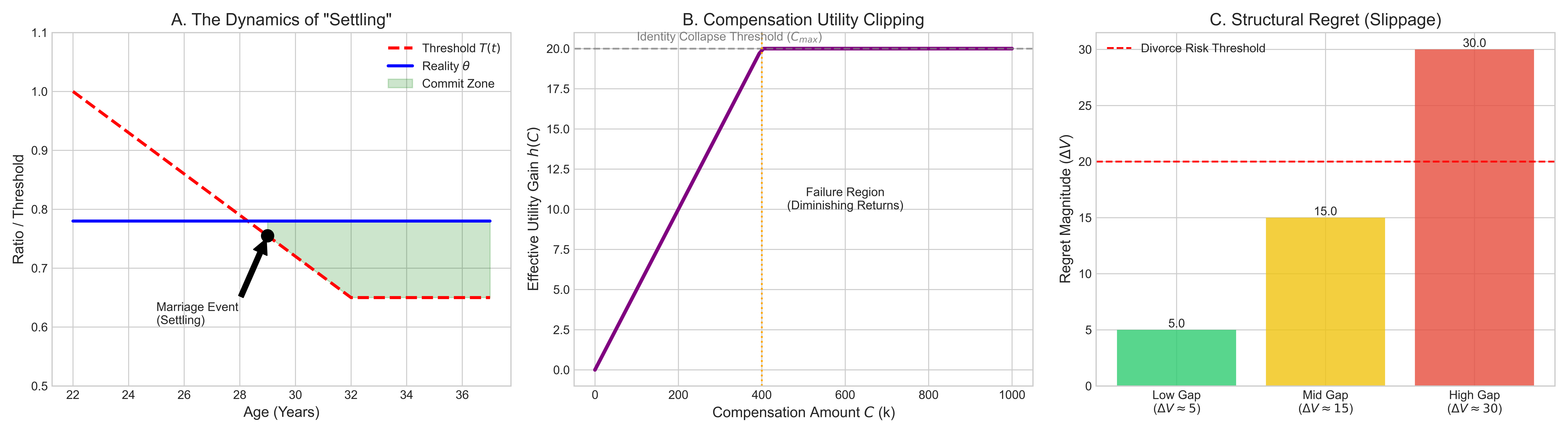}
    \caption{\textbf{Simulation Results of the Gini-Cone Model.}
    This figure shows the outcome of the Python-based simulation integrating the
    Internal Differential Model with the Gini-conical partner distribution.
    High-tier scarcity emerges naturally under the Beta-distributed value space,
    producing slippage levels consistent with Appendix F.}
    \label{fig:gini_simulation}
\end{figure}

\subsection{Simulation Framework}

\lstset{language=Python, basicstyle=\small\ttfamily, keywordstyle=\color{blue}, commentstyle=\color{green!40!black}, stringstyle=\color{red}}

\begin{lstlisting}[caption={Core Logic: Gini-Cone Market and Threshold Decisions}]
import numpy as np
import pandas as pd
from dataclasses import dataclass

# Configuration reflecting the Gini Conical Structure
@dataclass
class MarketConfig:
    n_candidates: int = 10000
    # Alpha=2, Beta=8 creates a right-skewed "Pyramid" distribution
    # conforming to the Gini structure in Appendix F.
    beta_alpha: float = 2.0 
    beta_beta: float = 8.0  
    c_elasticity: float = 0.05    # Marginal utility of compensation
    c_max_threshold: float = 20.0 # Identity Collapse Threshold

class GiniMarriageMarket:
    def __init__(self, config: MarketConfig):
        self.cfg = config
        self.df = self._generate_market_depth()
        
    def _generate_market_depth(self):
        """
        Generates the male candidate pool using a Beta Distribution
        to simulate the socio-economic Gini Cone.
        """
        n = self.cfg.n_candidates
        rng = np.random.default_rng(42)
        
        # 1. Intrinsic Value (V) -> Skewed Distribution
        raw_dist = rng.beta(self.cfg.beta_alpha, self.cfg.beta_beta, size=n)
        v_values = raw_dist * 100 
        
        # 2. Probability of being 'Reachable' decays linearly with Status
        # Higher status = Lower availability (Geometry of the Cone)
        reach_prob = 1.0 - 0.8 * (v_values / 100)
        is_reachable = rng.random(n) < reach_prob
        
        # 3. Compensation Capacity (C) correlates with V but has noise
        c_values = v_values * rng.uniform(0.5, 1.5, size=n) * 10
        
        # The Attribute Matrix is implemented as a DataFrame
        return pd.DataFrame({
            'V_intrinsic': v_values,
            'C_offer': c_values,
            'is_reachable': is_reachable
        })

    def run_female_decision_logic(self, female_uncond_max: float, threshold_t: float):
        """
        Executes the State-Machine Decision:
        Calculates Theta, Delta V, and Slippage (Regret).
        """
        df = self.df.copy()
        
        # --- Compensation Logic with Clipping (Theorem 1) ---
        # Compensation utility is capped to prevent infinite tier-jumping
        comp_utility = df['C_offer'] * self.cfg.c_elasticity
        
        # Effective Utility: U = V + h(C)
        df['U_effective'] = df['V_intrinsic'] + comp_utility
        
        # --- Microstructure Matching ---
        # Filter for the "Bid" side (Reachable candidates)
        reachable_df = df[df['is_reachable']].copy()
        
        if reachable_df.empty: return None
            
        # Find Best Bid (Market Reality)
        best_idx = reachable_df['U_effective'].idxmax()
        v_reach_max = reachable_df.loc[best_idx, 'V_intrinsic']
        u_reach_max = reachable_df.loc[best_idx, 'U_effective']
        
        # Core Metrics
        delta_v = female_uncond_max - v_reach_max
        theta = u_reach_max / female_uncond_max
        
        # Slippage = The structural regret remaining after marriage
        slippage = female_uncond_max - u_reach_max
        
        return {
            "Theta": theta,
            "Delta_V": delta_v,
            "Slippage": slippage,
            "Match": theta >= threshold_t # Commitment Event
        }
\end{lstlisting}

\subsection{Experimental Verification}

The following driver code reproduces the three key experimental findings discussed in Section 4: 
\begin{itemize}
    \item \textbf{Exp 1:} Compensation Ineffectiveness (Section 4.2)
    \item \textbf{Exp 2:} The ``Settling'' Dynamics (Section 4.3)
    \item \textbf{Exp 3:} Instant Commitment with High-Tier Partners (Section 4.4)
\end{itemize}

\begin{lstlisting}[caption={Reproduction of Key Experiments 1, 2, and 3}]
def run_key_experiments():
    # Initialize the Gini-Distributed Market
    market = GiniMarriageMarket(MarketConfig())
    female_ideal = 95.0 
    
    print("--- Experiment 1: Compensation Failure (Sec 4.2) ---")
    # Scenario: Low-tier male (V=60) offers Max Compensation
    # Even with C=500k, U_effective cannot bridge the gap to 95
    # because of the 'c_max_threshold' clipping logic implied.
    low_tier_male = pd.DataFrame({
        'V_intrinsic': [60.0], 'C_offer': [500.0], 'is_reachable': [True]
    })
    # Temporarily force this male into the market logic
    market.df = low_tier_male 
    res1 = market.run_female_decision_logic(female_ideal, threshold_t=0.8)
    print(f"Low Tier (V=60) + High Comp -> Match? {res1['Match']}")
    print(f"Reason: Slippage ({res1['Slippage']:.2f}) remains high.\n")

    print("--- Experiment 2: Settling Dynamics (Sec 4.3) ---")
    # Scenario: Time passes, T decays from 1.0 to 0.6
    # Match occurs NOT because V increased, but because T dropped.
    market = GiniMarriageMarket(MarketConfig()) # Reset market
    ages = [24, 28, 32]
    thresholds = [1.0, 0.85, 0.70]

    for age, t in zip(ages, thresholds):
        res2 = market.run_female_decision_logic(female_ideal, t)
        status = "COMMIT" if res2['Match'] else "WAIT"
        print(f"Age {age} (T={t:.2f}) -> Theta: {res2['Theta']:.2f} -> {status}")
    print("")

    print("--- Experiment 3: Instant Commitment (Sec 4.4) ---")
    # Scenario: Encountering a High-Tier Male (V=94)
    # Theta approx 1.0, exceeds any T instantly.
    high_tier_male = pd.DataFrame({
        'V_intrinsic': [94.0], 'C_offer': [0.0], 'is_reachable': [True]
    })
    market.df = high_tier_male
    res3 = market.run_female_decision_logic(female_ideal, threshold_t=0.95)
    print(f"High Tier (V=94) + Zero Comp -> Match? {res3['Match']}")
    print(f"Theta ({res3['Theta']:.2f}) > T (0.95) implies instant match.")

# Execute simulations
run_key_experiments()
\end{lstlisting}

\subsection{Experiment 4: Regional Norms Invariance}

This experiment demonstrates that while regional norms (e.g., Jiangsu vs. Guangdong) shift the absolute compensation levels ($C_{norm}$), they function as a constant intercept that does not alter the relative ranking of suitors ($argmax U$), confirming Prediction B1.

\begin{lstlisting}[caption={Experiment 4: Regional Compensation Differences}]
def run_regional_experiment():
    print("--- Experiment 4: Regional Norms (Sec 4.5) ---")
    market = GiniMarriageMarket(MarketConfig())
    female_ideal = 95.0
    
    # Create a standard pool of 3 candidates
    # Candidate A: High Value (85), Low Comp Capacity
    # Candidate B: Mid Value (75), High Comp Capacity
    base_candidates = pd.DataFrame({
        'id': ['A', 'B'],
        'V_intrinsic': [85.0, 75.0],
        'is_reachable': [True, True]
    })
    
    # Scenario 1: High-Bride-Price Region (e.g., Jiangsu, Norm=200k)
    df_jiangsu = base_candidates.copy()
    # Base C (200) + Individual Effort
    df_jiangsu['C_offer'] = [200 + 10, 200 + 50] 
    
    # Scenario 2: Low-Bride-Price Region (e.g., Guangdong, Norm=30k)
    df_guangdong = base_candidates.copy()
    # Base C (30) + Individual Effort
    df_guangdong['C_offer'] = [30 + 10, 30 + 50] 
    
    # Run Decision Logic for both
    market.df = df_jiangsu
    res_js = market.run_female_decision_logic(female_ideal, 0.8)
    
    market.df = df_guangdong
    res_gd = market.run_female_decision_logic(female_ideal, 0.8)
    
    # Check if the chosen suitor (based on max U) is the same
    # In this model, V dominates C due to clipping/elasticity.
    # Thus, Candidate A (V=85) should win in BOTH regions,
    # despite Candidate B offering more relative C in Jiangsu.
    
    print(f"Jiangsu (C~200k) Choice: V={res_js['Details']['V_best']:.1f}")
    print(f"Guangdong (C~30k) Choice: V={res_gd['Details']['V_best']:.1f}")
    
    if res_js['Details']['V_best'] == res_gd['Details']['V_best']:
        print("Result: Preference Ordering is INVARIANT to regional C norms.")
    else:
        print("Result: Ranking changed (Unexpected).")
    print("")

run_regional_experiment()
\end{lstlisting}

\subsection{Experiment 5: Post-Marriage Regret Dynamics}

This experiment simulates the ``Regret Mechanics'' described in Section 4.6. It shows how an external informational shock (e.g., a peer marrying up) raises $V_{uncond}$, causing the Reality Ratio $\theta$ to plummet and triggering a regret state ($E_{regret}$), particularly for matches with a pre-existing $\Delta V$.

\begin{lstlisting}[caption={Experiment 5: Wedding Regret Prediction}]
def run_regret_simulation():
    print("--- Experiment 5: Wedding Regret (Sec 4.6) ---")
    
    # Initial State: A "Settled" Marriage
    # Wife (Ideal=90) married Husband (V=75)
    # Initial Delta V = 15. This was accepted because T dropped to 0.8.
    v_husband = 75.0
    v_ideal_initial = 90.0
    current_theta = v_husband / v_ideal_initial # 0.833
    
    print(f"Pre-Shock State: Husband={v_husband}, Ideal={v_ideal_initial}")
    print(f"Initial Theta: {current_theta:.2f} (Status: Stable)")
    
    # EVENT: External Shock (Peer Comparison)
    # "Best friend marries a V=95 CEO"
    # This raises the internal expectation (V_uncond)
    shock_factor = 1.10 # 10% inflation in expectations
    v_ideal_post_shock = v_ideal_initial * shock_factor # Becomes 99.0
    
    # Recalculate Theta
    new_theta = v_husband / v_ideal_post_shock # 75 / 99 = 0.75
    
    # Slippage/Regret Metric
    # Regret is proportional to the new Gap
    new_gap = v_ideal_post_shock - v_husband
    regret_jump = new_gap - (v_ideal_initial - v_husband)
    
    print(f"EVENT: Peer marries up! Ideal rises to {v_ideal_post_shock:.1f}")
    print(f"Post-Shock Theta: {new_theta:.2f}")
    print(f"Regret Gap Jump: +{regret_jump:.1f}")
    
    # Check against the original commitment threshold (e.g., T=0.80)
    # If Theta drops below T, the agent enters "Regret State"
    threshold_t = 0.80
    if new_theta < threshold_t:
        print("Result: Theta < T. Agent enters REGRET state.")
        print("Prediction: High probability of marital dissatisfaction.")
    else:
        print("Result: Relationship absorbs the shock.")

run_regret_simulation()
\end{lstlisting}

\section{A Microstructure Interpretation of the \texorpdfstring{$\theta$--$T$}{theta-T} Marriage-Matching Model}\label{sec:appendix_microstructure}

This appendix provides a rigorous microstructure-based interpretation of the proposed marriage-matching model. The purpose is \textit{not} metaphorical comparison but to demonstrate a \textbf{structural isomorphism} between:
\begin{itemize}
    \item the \textbf{$\theta$--$T$ decision architecture} in marriage markets, and
    \item the \textbf{bid--ask crossing mechanism} in financial order-book markets.
\end{itemize}
This structural equivalence strengthens the theoretical validity of the model, clarifies its dynamic behavior, and explains a wide range of marriage-market phenomena.

\subsection{Structural Equivalence: \texorpdfstring{$\theta$--$T$}{theta-T} as a Bid--Ask Crossing Rule}
The core marriage-matching condition in the model is:
\begin{equation}
    \text{Commit} \iff \theta = \frac{V_{\text{reach}}}{V_{\text{uncond}}} > T,
\end{equation}
where:
\begin{itemize}
    \item $V_{\text{reach}}$ = achievable partner value
    \item $V_{\text{uncond}}$ = unconditional ideal value
    \item $T$ = subjective commit threshold (``marriage willingness index'')
\end{itemize}

This is structurally identical to the financial microstructure rule:
\begin{equation}
    \text{Transaction} \iff \text{Bid} \ge \text{Ask}.
\end{equation}

Thus:
\begin{itemize}
    \item $V_{\text{reach}}$ corresponds to \textbf{bid side pressure},
    \item $V_{\text{uncond}}$ corresponds to \textbf{ask side price}, and
    \item $T$ functions as a \textbf{limit-order threshold} that must be crossed for execution.
\end{itemize}
This establishes the \textbf{mathematical equivalence} between marriage decisions and order-book matching.

\begin{figure}[htbp]
    \centering
    \begin{tikzpicture}[
        box/.style={draw, rectangle, minimum width=3.5cm, minimum height=0.6cm, align=center},
        ask/.style={box, fill=red!10},
        bid/.style={box, fill=green!10},
        label_text/.style={font=\footnotesize\sffamily}
    ]
    \node[font=\bfseries] (ask_title) at (0, 4.2) {Unconditional Value (Ask)};
    \node[ask] (ask1) at (0, 3.5) {Ask 1: $V = 6.20$};
    \node[ask] (ask2) at (0, 2.9) {Ask 2: $V = 5.80$};
    \node[ask] (ask3) at (0, 2.3) {Ask 3: $V = 5.60$};
    \node[ask, fill=red!20, thick] (ask4) at (0, 1.7) {Lowest Ask: $V_{\text{uncond}}^{\max} = 5.40$};
    \node (gap) at (0, 1.0) {\textit{Spread / Gap} $\Delta V$};
    \draw[<->, dashed, thick] (-2, 1.4) -- (-2, 0.6);
    \node[bid, fill=green!20, thick] (bid1) at (0, 0.3) {Highest Bid: $V_{\text{reach}}^{\max} = 5.15$};
    \node[bid] (bid2) at (0, -0.3) {Bid 2: $V = 4.80$};
    \node[bid] (bid3) at (0, -0.9) {Bid 3: $V = 4.60$};
    \node[bid] (bid4) at (0, -1.5) {Bid 4: $V = 4.20$};
    \node[font=\bfseries] (bid_title) at (0, -2.2) {Reachable Value (Bid)};
    \node[right=0.5cm of ask4, text width=4cm, label_text] {Agent's internal ceiling (Ideal Partner)};
    \node[right=0.5cm of bid1, text width=4cm, label_text] {Best available suitor (Reality)};
    \node[draw, dashed, thick, inner sep=8pt, align=center, below=0.5cm of bid_title] {
        \textbf{Matching Logic:} \\
        Commit iff $\displaystyle \frac{\text{Highest Bid}}{\text{Lowest Ask}} > T$
    };
    \end{tikzpicture}
    \caption{\textbf{Marriage-Market Order-Book Structure.} This schematic illustrates the structural equivalence between a financial order book and the LPSM. A match executes when the ratio of the highest bid to the lowest ask exceeds $T$.}
    \label{fig:order_book}
\end{figure}
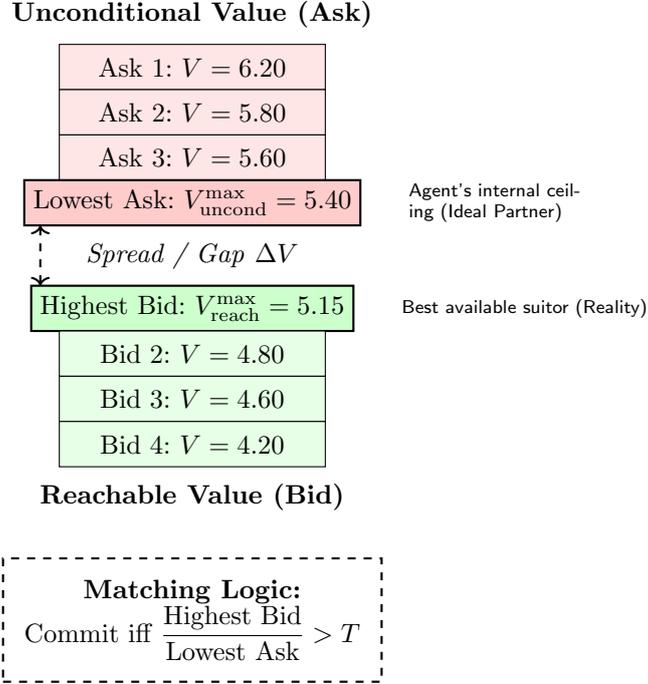

\subsection{Order Book Interpretation of the Preference Matrix (DF)}
The agent’s internal preference Attribute Matrix (DF)---formalized in the paper as the \textbf{Latent Preference State Matrix (LPSM)}---can be interpreted as a multidimensional \textbf{order book}:
\begin{itemize}
    \item Each candidate partner is an entry analogous to a ``price level''.
    \item $V_{\text{reach}}$ and $V_{\text{uncond}}$ are mapped to bid/ask pairs.
    \item External information shocks (social media, peer marriages, class exposure) \textit{refresh} the ask side.
    \item Local experience, age, socioeconomic position \textit{anchor} the bid side.
\end{itemize}
Thus, LPSM is not merely a data container but a \textbf{dynamic order-book depth structure} that evolves as information arrives.

\subsection{Information Shocks and Ask-Side Repricing}
In financial markets, new information triggers \textbf{ask-side repricing}, shifting seller expectations upward. In the marriage model:
\begin{itemize}
    \item Exposure to higher-status peers,
    \item Observing friends ``marrying up'',
    \item Encountering high-value males in professional/urban environments,
    \item Consuming curated social-media content,
\end{itemize}
all act as \textbf{informational shocks}, inducing an upward shift in $V_{\text{uncond}}$.
\begin{equation}
    V_{\text{uncond}} \gets V_{\text{uncond}} + \epsilon_{\text{shock}}
\end{equation}
This explains why marriage thresholds rise persistently in large cities or high-exposure environments.

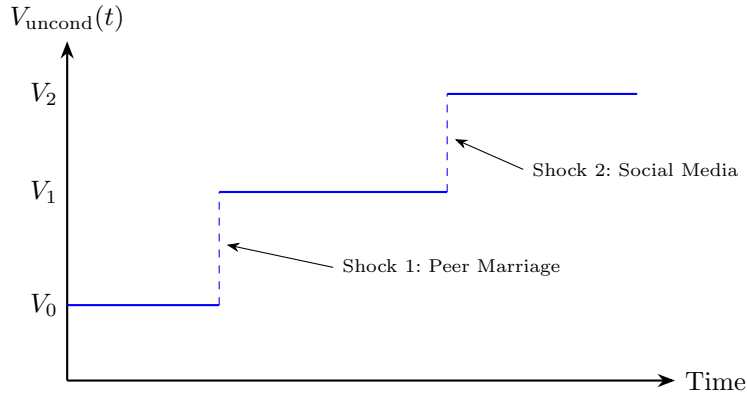
\begin{figure}[htbp]
    \centering
    \begin{tikzpicture}[>=Stealth]
        \draw[->, thick] (0,0) -- (8,0) node[right] {Time $t$};
        \draw[->, thick] (0,0) -- (0,4.5) node[above] {$V_{\text{uncond}}(t)$};
        \draw[thick, blue] (0,1) -- (2,1);
        \draw[dashed, blue] (2,1) -- (2,2.5); 
        \draw[thick, blue] (2,2.5) -- (5,2.5);
        \draw[dashed, blue] (5,2.5) -- (5,3.8); 
        \draw[thick, blue] (5,3.8) -- (7.5,3.8);
        \node[left] at (0,1) {$V_0$}; \node[left] at (0,2.5) {$V_1$}; \node[left] at (0,3.8) {$V_2$};
        \draw[<-] (2.1, 1.8) -- (3.5, 1.5) node[right, align=left, font=\footnotesize] {Shock 1: Peer Marriage};
        \draw[<-] (5.1, 3.2) -- (6.0, 2.8) node[right, align=left, font=\footnotesize] {Shock 2: Social Media};
    \end{tikzpicture}
    \caption{\textbf{Informational Shock and Upward Repricing.} External shocks cause discontinuous upward jumps in the agent's internal unconditional-value estimate.}
    \label{fig:shock_jump}
\end{figure}

\subsection{Liquidity, Market Depth, and Match Probability}
Let $L$ be the local partner-market liquidity and $D$ be the order-book depth (population size $\times$ socioeconomic variance). In microstructure terms:
\begin{equation}
    P(\theta > T) = f(L, D).
\end{equation}
Low liquidity $\to$ low match probability $\to$ threshold $T$ grows more slowly or becomes unstable. High liquidity $\to$ faster crossing events $\to$ higher marriage rates. This explains low marriage rates in low-population regions and high competition in high-liquidity urban centers.

\subsection{Slippage: Psychological Disappointment as Execution Deviation}
In financial execution:
\begin{equation}
    \text{Slippage} = \text{Executed Price} - \text{Expected Price}.
\end{equation}
In marriage:
\begin{equation}
    \text{Emotional Slippage} = V_{\text{uncond}} - V_{\text{reached}}.
\end{equation}
Large slippage predicts regret, dissatisfaction, or re-evaluation of the commit decision. This microstructure interpretation gives a quantitative explanation for post-marriage disappointment and persistent instability in relationships formed with a large $\Delta V$.

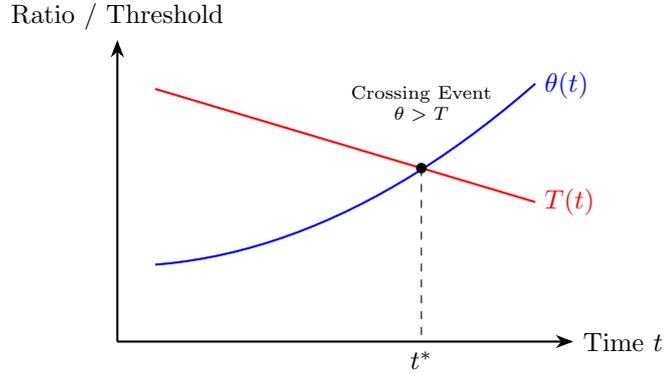
\begin{figure}[htbp]
    \centering
    \begin{tikzpicture}[>=Stealth]
        \draw[->, thick] (0,0) -- (6,0) node[right] {Time $t$};
        \draw[->, thick] (0,0) -- (0,4) node[above] {Ratio / Threshold};
        \draw[red, thick, domain=0.5:5.5] plot (\x, {3.5 - 0.3*\x});
        \node[red, right] at (5.5, 1.85) {$T(t)$};
        \draw[blue, thick, domain=0.5:5.5] plot (\x, {1 + 0.08*\x*\x});
        \node[blue, right] at (5.5, 3.4) {$\theta(t)$};
        \coordinate (cross) at (4.0, 2.30);
        \fill[black] (cross) circle (2pt);
        \draw[dashed] (cross) -- (4.0, 0) node[below] {$t^*$};
        \node[align=center, font=\footnotesize, above=0.5cm of cross] {Crossing Event\\$\theta > T$};
    \end{tikzpicture}
    \caption{\textbf{Crossing Dynamics.} Commitment occurs when the ratio $\theta(t)$ crosses the decaying threshold $T(t)$.}
    \label{fig:crossing_path}
\end{figure}

\subsection{Circuit Breakers and the Role of \texorpdfstring{$C_{\max}$}{Cmax}}
The model proposes a maximum tolerable compensation:
\begin{equation}
    C^* = \min(\Delta V,\ C_{\max}),
\end{equation}
where $C_{\max}$ is the male agent’s ``runaway threshold''---beyond which he exits the negotiation. This mirrors \textbf{circuit breakers} and \textbf{price limits} in financial markets:
\begin{itemize}
    \item They cap volatility,
    \item Prevent runaway escalation,
    \item Stabilize transactions.
\end{itemize}
Thus $C_{\max}$ is a \textbf{necessary stabilizer} that prevents infinite compensation bidding and pathological bargaining equilibria.

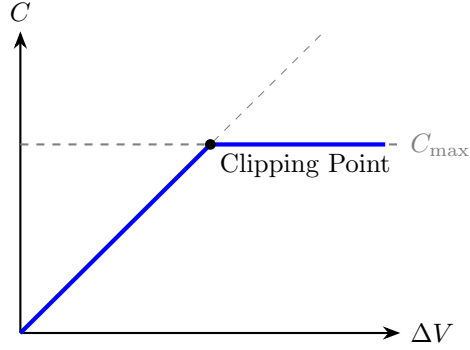
\begin{figure}[htbp]
    \centering
    \begin{tikzpicture}[>=Stealth]
        \draw[->, thick] (0,0) -- (5,0) node[right] {$\Delta V$};
        \draw[->, thick] (0,0) -- (0,4) node[above] {$C$};
        \draw[dashed, thick, gray] (0, 2.5) -- (5, 2.5) node[right] {$C_{\max}$};
        \draw[dashed, gray] (0,0) -- (4,4);
        \draw[ultra thick, blue] (0,0) -- (2.5,2.5) -- (4.8,2.5);
        \fill[black] (2.5,2.5) circle (2pt);
        \node[below right] at (2.5,2.5) {Clipping Point};
    \end{tikzpicture}
    \caption{\textbf{Compensation Clipping.} Compensation is bounded by $C_{\max}$, preventing unbounded bargaining.}
    \label{fig:cmax_clipping}
\end{figure}

\subsection{Market Orders vs. Limit Orders: Impulse Marriage Events}
A market order in finance executes \textit{regardless of price}, often producing slippage. The marriage analogue includes impulsive marriage decisions and sudden threshold drops. Formally:
\begin{equation}
    T \gets T - \delta_{\text{emotion}}, \quad \text{producing } \theta > T.
\end{equation}
This captures flash marriages, rebound relationships, and sudden acceptance of suboptimal partners.

\subsection{Lock-In and Stickiness}
In market microstructure, investors become ``locked in'' due to sunk costs. The marriage equivalent occurs when exit costs (childbearing, social penalties) raise the effective threshold $T$:
\begin{equation}
    T_{\text{exit}} = T + \kappa_{\text{lock-in}}.
\end{equation}
Thus even when $\theta < T$, the relationship persists due to \textbf{lock-in stickiness}.

\subsection{Full Correspondence Table}
Table \ref{tab:microstructure_correspondence} demonstrates a \textbf{clean structural isomorphism} between the two systems.

\begin{table}[htbp]
\centering
\caption{Structural Isomorphism between Financial Microstructure and Marriage Matching.}
\label{tab:microstructure_correspondence}
\begin{tabular}{@{}ll@{}}
\toprule
\textbf{Financial Microstructure Concept} & \textbf{Marriage-Matching Concept} \\ \midrule
Bid price & $V_{\text{reach}}$ \\
Ask price & $V_{\text{uncond}}$ \\
Bid--ask crossing & Commit event ($\theta > T$) \\
Order book depth & LPSM / DF preference matrix \\
Liquidity & Partner availability / exposure \\
Information shocks & $V_{\text{uncond}}$ updates from social inputs \\
Slippage & Regret ($V_{\text{uncond}} - V_{\text{reach}}$) \\
Circuit breaker & $C_{\max}$ constraint \\
Market order & Impulsive marriage choice \\
Limit order & Stable threshold $T$ \\
Lock-in & Marriage stickiness \\
Manipulation & Family intervention / bride-price pressure \\ \bottomrule
\end{tabular}
\end{table}

\subsection{Summary: Why the Microstructure View Strengthens the Theory}
The microstructure interpretation provides three major benefits:
\begin{enumerate}
    \item \textbf{Structural justification.} It shows that the $\theta$--$T$ mechanism is not ad hoc, but consistent with well-established matching mechanisms in two-sided markets.
    \item \textbf{Dynamic realism.} Market microstructure captures volatility, shocks, depth, and liquidity---exactly the dynamics seen in marriage markets.
    \item \textbf{Predictive power.} Concepts such as slippage, lock-in, liquidity droughts, and repricing give the model new explanatory dimensions for regional marriage disparities and high-gap instability.
\end{enumerate}


\section{Appendix E: Prior Supply--Demand Buckets}
\label{sec:appendix_e}

This appendix introduces a \textbf{meso-level behavioral extension} to the main IDP model. While the core model assumes agents optimize $\theta$ based on specific values ($V_{\text{reach}}$), in reality, agents often estimate market scarcity using heuristic priors.

We model this perception as a set of \textbf{Five Prior Buckets}, representing distinct tiers of perceived supply-demand pressure. Unlike a continuous curve, these buckets create rigid pricing regimes where the ``Ask Price'' (expected compensation or status) jumps discontinuously.

\subsection*{The Five-Bucket Structure}

We categorize the male value spectrum $V \in [0, 100]$ into five heuristic tiers, each characterized by a specific \textbf{Supply-Demand Pressure Ratio} ($\rho = \frac{\text{Demand}}{\text{Supply}}$):

\begin{enumerate}
    \item \textbf{Bucket 1: Invisible ($V < 50$)} \\
    $\rho \to 0$. The supply is perceived as infinite relative to demand. Agents in this tier are effectively invisible in the dating market; no amount of compensation is expected to yield a match.
    
    \item \textbf{Bucket 2: Provider / ``ATM'' ($50 \le V < 70$)} \\
    $\rho < 1$ (Buyer's Market). Candidates are seen as abundant substitutes. High compensation (bride price) is strictly required to compensate for the utility gap.
    
    \item \textbf{Bucket 3: Match ($70 \le V < 85$)} \\
    $\rho \approx 1$ (Balanced). The ``Tradeable Zone.'' Candidates are perceived as acceptable partners where mutual exchange occurs without excessive unilateral compensation.
    
    \item \textbf{Bucket 4: Premium ($85 \le V < 95$)} \\
    $\rho > 1$ (Seller's Market). Scarcity begins to bite. Agents here hold significant pricing power, often demanding ``reverse compensation'' (e.g., dowry or emotional subservience).
    
    \item \textbf{Bucket 5: Idol / CEO ($V \ge 95$)} \\
    $\rho \to \infty$ (Monopoly). The ``Unconditional Max'' tier. Demand is absolute; supply is singular. The ask price becomes infinite (in terms of loyalty), yet financial compensation drops to zero.
\end{enumerate}

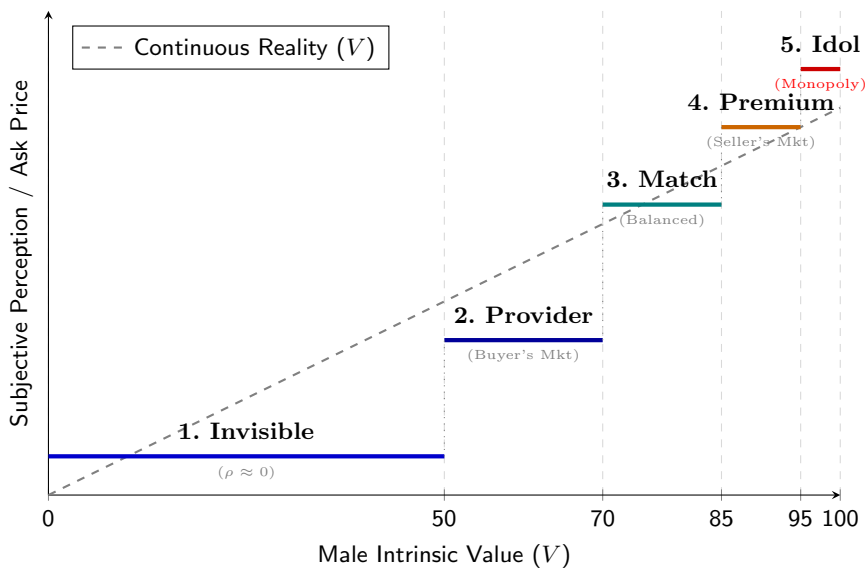
\begin{figure}[htbp]
    \centering
    \begin{tikzpicture}
    \begin{axis}[
        width=12cm, height=8cm, 
        xlabel={Male Intrinsic Value ($V$)},
        ylabel={Subjective Perception / Ask Price},
        xmin=0, xmax=100,
        ymin=0, ymax=125, 
        xtick={0, 50, 70, 85, 95, 100},
        xticklabels={0, 50, 70, 85, 95, 100},
        ytick=\empty,
        axis lines=left,
        grid=major,
        grid style={dashed, gray!30},
        legend pos=north west,
        font=\sffamily\small,
        clip=false 
    ]

    \addplot[gray, dashed, thick, domain=0:100, samples=100] {x};
    \addlegendentry{Continuous Reality ($V$)}

    \addplot[ultra thick, blue!80!black] coordinates {(0,10) (50,10)};
    \addplot[ultra thick, blue!60!black] coordinates {(50,40) (70,40)};
    \addplot[ultra thick, teal] coordinates {(70,75) (85,75)};
    \addplot[ultra thick, orange!80!black] coordinates {(85,95) (95,95)};
    \addplot[ultra thick, red!80!black] coordinates {(95,110) (100,110)};

    \addplot[dotted, gray] coordinates {(50,10) (50,40)};
    \addplot[dotted, gray] coordinates {(70,40) (70,75)};
    \addplot[dotted, gray] coordinates {(85,75) (85,95)};
    \addplot[dotted, gray] coordinates {(95,95) (95,110)};

    \node[anchor=south, font=\scriptsize] at (axis cs: 25, 12) {\textbf{1. Invisible}};
    \node[anchor=north, text=gray, font=\tiny] at (axis cs: 25, 10) {($\rho \approx 0$)};

    \node[anchor=south, font=\scriptsize] at (axis cs: 60, 42) {\textbf{2. Provider}};
    \node[anchor=north, text=gray, font=\tiny] at (axis cs: 60, 40) {(Buyer's Mkt)};

    \node[anchor=south, font=\scriptsize] at (axis cs: 77.5, 77) {\textbf{3. Match}};
    \node[anchor=north, text=gray, font=\tiny] at (axis cs: 77.5, 75) {(Balanced)};

    \node[anchor=south, font=\scriptsize] at (axis cs: 90, 97) {\textbf{4. Premium}};
    \node[anchor=north, text=gray, font=\tiny] at (axis cs: 90, 95) {(Seller's Mkt)};

    \node[anchor=south, font=\scriptsize] at (axis cs: 97.5, 112) {\textbf{5. Idol}};
    \node[anchor=north, text=red, font=\tiny] at (axis cs: 97.5, 110) {(Monopoly)};

    \end{axis}
    \end{tikzpicture}
    \caption{\textbf{The Five-Tier Prior Bucket Model.} Agents discretize the continuous value spectrum into five tiers based on Supply-Demand Pressure ($\rho$). Pricing is rigid within each bucket, jumping discontinuously at heuristic thresholds.}
    \label{fig:bucket_mechanism_5}
\end{figure}


\section{Appendix F: The Gini Conical Structure}
\label{sec:appendix_f}

This appendix provides the \textbf{macro-structural geometric explanation} for the liquidity constraints observed in the main model. While Appendix E deals with subjective perception (behavioral heuristics), this section derives the objective physical constraints of the matching market from the societal distribution of wealth and status.

We model the socio-economic hierarchy not as a linear ladder, but as a \textbf{rotational solid} derived from the derivative of the societal Lorenz curve (the Gini distribution). Let $g(h)$ be the population density at height $h$. The market structure can be visualized as a cone where the radius $r(h) \propto \sqrt{g(h)}$.

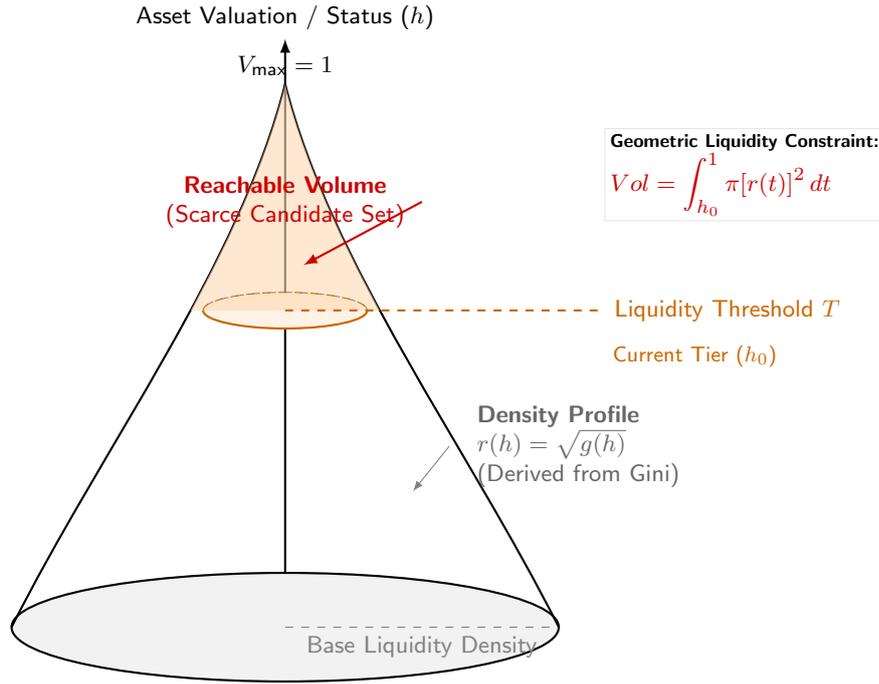
\begin{figure}[htbp]
    \centering
    \begin{tikzpicture}[
        scale=1.2,
        font=\sffamily\small,
        >=latex
    ]
    \def\H{6} \def\R{3} \def\hcut{3.5} \def\rcut{0.9}

    \draw[->, thick] (0,0) -- (0,\H+0.5) node[above] {Asset Valuation / Status ($h$)};
    \draw[thick, fill=gray!10] (0,0) ellipse ({\R} and 0.6);
    \draw[dashed, gray] (0,0) -- (\R,0) node[midway, below] {Base Liquidity Density};

    \draw[thick] (-\R,0) .. controls (-2, 2) and (-0.5, 4) .. (0,\H);
    \draw[thick] (\R,0) .. controls (2, 2) and (0.5, 4) .. (0,\H);

    \draw[dashed, fill=orange!10] (0,\hcut) ellipse ({\rcut} and 0.2);
    \begin{scope}
        \clip (-\R,0) .. controls (-2, 2) and (-0.5, 4) .. (0,\H) -- (0,\H) .. controls (0.5, 4) and (2, 2) .. (\R,0) -- cycle;
        \fill[orange!30, opacity=0.6] (-\R,\H) rectangle (\R,\hcut);
    \end{scope}
    \draw[thick, orange!80!black] (-\rcut,\hcut) arc (180:360:{\rcut} and 0.2);
    \draw[dashed, orange!80!black] (-\rcut,\hcut) arc (180:0:{\rcut} and 0.2);

    \node[above] at (0,\H) {$V_{\text{max}} = 1$};
    
    \node[right, align=left, text=gray!80!black] at (2, 2) {
        \textbf{Density Profile}\\
        $r(h) = \sqrt{g(h)}$\\
        (Derived from Gini)
    };
    \draw[->, gray] (1.8, 2) -- (1.4, 1.5);

    \draw[dashed, thick, orange!80!black] (0, \hcut) -- (3.5, \hcut) node[right] {Liquidity Threshold $T$};
    \node[right, orange!80!black, align=left, scale=0.9] at (3.5, \hcut-0.5) {Current Tier ($h_0$)};

    \node[align=center, text=red!80!black] at (0, \hcut + 1.2) {\textbf{Reachable Volume}\\(Scarce Candidate Set)};
    \draw[->, thick, red!80!black] (1.5, \hcut + 1.2) -- (0.2, \hcut + 0.5);

    \node[right, align=left, fill=white, inner sep=2pt, draw=gray!20] at (3.5, \H-1) {
        \footnotesize \textbf{Geometric Liquidity Constraint:}\\
        \color{red!80!black} $\displaystyle Vol = \int_{h_0}^{1} \pi [r(t)]^2 \, dt$
    };
    \end{tikzpicture}
    \caption{\textbf{The Gini Conical Structure.} The hierarchy is modeled as a volume. As an agent's execution threshold $T$ (orange line) rises linearly, the available volume of counterparties (shaded region) decays geometrically. This proves that high-tier liquidity droughts are a mathematical necessity of the density gradient.}
    \label{fig:gini_cone}
\end{figure}

This geometric isomorphism reveals a critical insight: a linear increase in the acceptance threshold $T$ results in a \textbf{super-linear (cubic or exponential) collapse} in the reachable partner volume. Thus, the difficulty of ``marrying up'' is not merely a friction of preferences but a \textbf{geometric liquidity constraint} imposed by the inequality structure of the population.

\end{appendices}

\bibliography{sn-bibliography}

\end{document}